\documentstyle[12pt]{article}
\newcommand{\be}{\begin{equation}}
\newcommand{\ee}{\end{equation}}
\newcommand{\rref}[1]{(\ref{#1})}

\input epsf

\begin{document}

\begin{flushright}
ULB-TH-97/04\\
PAR-LPTHE-97-08\\
\end{flushright}
~\\

\begin{center}
{\Large A note on the gauge symmetries of pure Chern-Simons theories
with $p$-form gauge fields} \\

~\\
~\\

M. Ba\~{n}ados$^{1,2}$, M. Henneaux$^{1,3}$, C. Iannuzzo$^3$
and C.M. Viallet$^4$
\end{center}

\begin{center}
{\it $^1$Centro de Estudios
Cient\'{\i}ficos de Santiago, Casilla 16443, Santiago, Chile\\
$^2$Departamento de  F\'{\i}sica, Universidad de Santiago,
Casilla 307, Correo 2, Santiago, Chile\\
$^3$Facult\'e des Sciences, Universit\'e Libre de Bruxelles, Campus
Plaine, \\ C.P. 231, B-1050, Bruxelles, Belgium\\
$^4$Centre National de la Recherche Scientifique \\
Laboratoire de Physique Th\'eorique et Hautes Energies, \\
Universit\'es Paris VI et Paris VII, Bte 126, 4, Place Jussieu,\\
75252 Paris Cedex 05, France}

\end{center}

\begin{abstract}
The gauge symmetries of pure Chern-Simons theories with $p$-form
gauge fields are analyzed. It is shown that the number of independent
gauge symmetries depends crucially on the parity of $p$. 
The case where $p$ is odd appears to be a direct generalization of
the $p=1$ case and presents the
remarkable feature that the timelike diffeomorphisms can be expressed
in terms of the spatial diffeomorphisms and the internal gauge
symmetries. By constrast, the timelike diffeomorphisms may be
an independent  gauge symmetry when $p$ is even.  This happens when
the number of fields and the spacetime dimension fulfills an
algebraic condition which is explicitely written.

\end{abstract}

\vfill
\break

\section{Introduction}
\setcounter{equation}{0}

Pure Chern-Simons theory in $3$ dimensions is one of the most studied
examples
of a topological field theory. It is a model which does not involve
the spacetime metric but is yet generally covariant.
Furthermore, as noted by many authors \cite{Jackiw}, 
the spacetime diffeomorphisms
\be
\delta_{\xi} A^{a}_{\mu} = \pounds_{\xi} A^{a}_{\mu} =
  \xi ^{\rho} F^{a}_{\rho \mu} + D_{\mu} (\xi ^{\rho} A^{a}_{\rho }),
\label{1.1}
\ee
are not independent from the internal gauge symmetries
\be \delta_{\xi} A^{a}_{\mu} = D_{\mu} \epsilon^{a}
\label{1.2}
\ee
since \rref{1.1} reduces to \rref{1.2} with $ \epsilon^{a} =
\xi^{\rho} A^{a}_{\rho}$ when the equations of motion hold ($
F^{a}_{\rho \mu} = 0$). In \rref{1.1} and \rref{1.2}, the $A^{a}_{\mu
}$ are the components of the connection, while the $F^{a}_{\rho \mu}$
are those of the curvature $2$-form, respectively.

Due to the non-independence of the spacetime diffeomorphisms, the
only first class-constraints in the Hamiltonian formulation are the
constraints associated with the internal gauge symmetry \rref{1.2}.
There is no independent first class constraint associated with
\rref{1.1}. Thus, by finding the most \hyphenation{general} general
state invariant under \rref{1.2}, one may automatically produce
states which are invariant under \rref{1.1}.
Implementing diffeomorphism invariance is then a mere
byproduct  of implementing invariance under internal
gauge symmetries. 

It was shown in recent publications \cite{B.G.H.} that this crucial
property of three-dimensional Chern-Simons theory does not survive in
the higher dimensional generalizations. In those models, described by
the Lagrangian $(2k+1)$-form
\be 
{\cal L }^{(2k+1)}_{C.S.} = g_{a_{1}\dots a_{k+1}} F^{a_1}\wedge
\dots
  \wedge F^{a_k} \wedge A^{a_{k+1}} + ``more" \\ (k\geq 2)
\label{1.3}
\ee where $g_{a_{1}\dots a_{k+1}}$ is an invariant tensor and where
$``more"$ completes \rref{1.3} so as to make it invariant under
\rref{1.2} up to an exact term. It is generically no longer true that
the diffeomorphisms \rref{1.1}, which are still symmetries of
\rref{1.3}, can be expressed in terms of the internal gauge
transformations \rref{1.2}.  The reason is that the equations of
motion no longer imply the vanishing of the curvature two-form
$F^{a}_{\mu \nu }$. At the same time, the theory described by
(\ref{1.3}) does possess degrees of freedom for $k>1$ while it does
not for $k=1$.

However, the higher-dimensional Chern-Simons models still possess a
remarkable feature. Namely, not all the spacetime diffeomorphisms are
independent gauge transformations, but only those defined by {\it
spatial} 
reparameterizations. As established in \cite{B.G.H.}, the timelike
diffeomorphisms can be expressed in terms of the internal gauge
symmetries and the spacelike diffeomorphisms. This implies that, in
the Hamiltonian formalism, there are first class constraints $G_{a}
\approx 0$ generating the internal gauge symmetries \rref{1.2} and
first class constraints ${\cal H}_{i} \approx 0$ generating the
spatial diffeomorphisms, but there are no independent first class
constraints associated with the timelike diffeomorphisms. In order to
find the most general gauge invariant state, it suffices to solve the
conditions ${\cal H}_{i} |\psi >=0$ and $G_{a} |\psi > = 0$. Since
these conditions are purely kinematical, they are in principle more
easily amenable to an exact resolution (using e.g. the loop
representation \cite{loop} ). From this point of view, Chern-Simons
theories in higher dimensions are of great interest to relativists
since they provide non trivial models (models with a finite 
number of local degrees of freedom per space point) in which the problem
of
implementing quantum-mechanically invariance under the full
diffeomorphism group is reduced to the kinematical problem of finding
quantum states invariant under spatial diffeomorphisms. They are in
that respect comparable to the toy models analyzed in \cite{HK}. 

In the search for other generally covariant theories with the same
property, we have analyzed the dynamics of pure Chern-Simons theories
with forms of higher degree. These are described by the Lagrangian 
$n$-form
\be 
{\cal L} = g_{a_1 \dots a_{k+1}}  B^{a_1} \wedge H^{a_2} \wedge \dots
\wedge H^{a_{k+1}} 
\label{1.4} 
\ee
where $B^{a} (a = 1,\dots ,N)$ are $p$-forms and $H^{a}$
are the corresponding (Abelian) curvatures,
\be  
H^a = dB^a.
\label{1.5}
\ee The spacetime dimension $n$ is equal to $p+k(p+1)$. We shall call
$k$ the order of the Chern-Simons theory.  The ($p+1$)-forms $H^a$ are
commuting when $p$ is odd and anticommuting when $p$ is even.  If $p$
is odd, the not totally symmetric parts of $g_{a_1 \dots a_{k+1}}$
contribute a total derivative to ${\cal L}$, and one may then assume that
$g_{a_1 \dots a_{k+1}}$ is totally symmetric.  Similarly, if $p$ is
even, we will assume that $g_{a_1 \dots a_{k+1}}$ is totally
antisymmetric.  The Chern-Simons Lagrangian is such that in $(n+1)$
dimensions, $ \int d{\cal L}$ is the topological invariant $\int
g_{a_1 \dots a_{k+1}} H^{a_1} \wedge \dots \wedge H^{a_{k+1}}$. The
Lagrangian ${\cal L}$ is invariant under the usual $p$-form Abelian
gauge transformations \be \delta _{\Lambda } B^{a} = d\Lambda ^{a}
\label{1.6}
\ee
where $\Lambda ^{a}$ are $N$ $(p-1)$-forms, as well as under
spacetime
diffeomorphisms
\be
\delta _{\xi } B^{a} = \pounds_{\xi } B^{a} \mbox{.}
\label{1.7}
\ee
The first gauge symmetry \rref{1.6} is reducible since if one takes
$\Lambda ^{a} = d\mu ^{a}$, one gets $\delta _{\Lambda } B^{a} = 0$
(identically). The second gauge symmetry \rref{1.7} can be rewritten
equivalently as
\be \delta ^{\prime }_{\xi } B^a = i_{\xi } H^a
\label{1.8}
\ee
since $\pounds_{\xi } = i_{\xi } d + d i_{\xi }$. 

The equations of motion following from \rref{1.4} are explicitly
\be
g_{a_1 a_2 \dots a_{k+1}} H^{a_2} \wedge \dots \wedge H^{a_{k+1}}
= 0
\label{1.9}
\ee
and do not imply that the curvatures $H^{a}$ vanish, unless the order
$k$ is
equal to one.

$P$-form gauge fields appear systematically in supergravity theories
in higher dimensions \cite{sugra}. A Chern-Simons-like term appears
in the Lagrangian of $n=11$ supergravity \cite{julia}, besides the
kinetic term proportional to $H^{\mu\nu\lambda\rho}
H_{\mu\nu\lambda\rho}$.

We have found that the pure Chern-Simons theories with Lagrangian
density \rref{1.4} generically have local degrees of freedom when $k
\geq 2$. This is not surprising in view of the analysis of
\cite{B.G.H.}. The new result, however, is that the timelike
diffeomorphisms {\it may now be independent gauge symmetries}. 

This will occur only when $p$ is even, and the number $\kappa$ defined by
\be \kappa \equiv N \left(
\begin{array}{c} n-1 \\ p \end{array} \right) -(n-1)
\ee
is an {\underline odd integer}.

The number $\kappa$ is the difference between the total number of spatial
components of  the set of $p$-forms, and the number of spatial dimensions
in the theory.

The above condition can be fulfilled by choosing appropriately $N$ and $k$
(for a given even $p$). There is thus a striking difference between
pure Chern-Simons theories with $p$-forms of even degree and pure
Chern-Simons theories with $p$-forms of odd degree. This difference
could not have been anticipated by a mere look at the gauge symmetries
\rref{1.6},\rref{1.7},\rref{1.8} - which take the same form in all
cases - and requires a more detailed investigation of the dynamics.
Our results imply that the dependence of the timelike diffeomorphisms
on the other gauge symmetries found for $p=1$ in \cite{B.G.H.} is not
a remnant of the topological construction leading to these models.
Indeed this construction is identical for all values of $p$, $N$ or
$k$. The phenomenon has a different origin which is yet to be
uncovered.

\section{The case $p=2$, $k=2$}

\subsection{Constraints} \setcounter{equation}{0}

We begin the discussion with $p=2$. The condition $\kappa$= odd integer
becomes, using $n =k(p+1) + p$:
\be  
(3k+1) [\frac{3kN}{2} - 1 ] = \mbox{odd integer}
\label{2.1}
\ee
and cannot be realized for $k=1$ for which the theory has actually no
degrees of freedom. We thus take the simplest case, namely $k=2$,
which
yields
\be 7[3N - 1] = \mbox{odd integer,}
\label{2.2}
\ee
a condition that is fulfilled if and only if $N$ is an even integer.

The goal of this section is to show that indeed, the timelike
diffeomorphisms can be expressed in terms of the internal gauge
transformations and the spatial diffeomorphisms for \underline{odd}
values of $N$, while they are independent gauge symmetries for
\underline{even} values of $N$ (except for low values of $N$ ($N<8$)
where there are accidental degeneracies\footnote{For $N=1$ and $N=2$,
the Lagrangian vanishes identically and those theories are then
trivial. For $N=4$, one field identically drops out because any
antisymmetric tensor $g_{abc}$ admits a ``zero vector" $\lambda^a\neq
0$, solution of the equation $g_{abc} \lambda^c=0$. So, one is
effectively reduced to the $N=3$ case plus one field with zero
Lagrangian. Finally, the case $N=6$ appears to be equivalent to two
independent $N=3$ theories and thus has twice the number of gauge
symmetries as the $N=3$ case.}).  
The proof of this property turns out
to be of mathematical interest in its own right since it involves the
discrete projective plane with $7$ points related to the octonion
algebra.

In order to analyze the gauge symmetries, we rewrite the Lagrangian
\rref{1.4} in Hamiltonian form. It is well appreciated by now that
the concept of gauge symmetry cannot be thought of independently from
the dynamics. This appears quite strikingly when discussing the
(in)dependence of a given set of a gauge transformations, since the
form of the ``on-shell trivial" gauge symmetries explicitly involves
the dynamics. It is for this reason that the Hamiltonian formulation,
where the dynamics takes a transparent form, is convenient in the
analysis of the gauge symmetries - see for instance \cite{HT}. With
$p=2$ and $k=2$, the
Lagrangian \rref{1.4} reads
\be 
{\cal L} = g_{abc} B^a \wedge H^b \wedge H^c
\label{2.3}
\ee
where $H^b $ is the $3$-form $dB^a$, and where $g_{abc}$ is
completely antisymmetric.

When split into space and time, the action (\ref{2.3}) is up to a
boundary term equal to
\be
S = \int dx^0 d^7 x [ l^{ij}_a  \dot{B} ^a_{ij} - B^a_{0i} K^i
_a ]
\label{2.4}
\ee
with
\be
l^{ij} _a = \epsilon ^{ijk_1 \dots k_5} g_{abc} B^b _{k_1 k_2 }
H^c _{k_3 k_4 k_5}
\label{2.5}
\ee
and
\be 
K^i _a = \epsilon ^{ij_1 j_2 j_3 j_4 j_5 j_6} g_{abc} H^b _{j_1
j_2 j_3}
H^c _{j_4 j_5 j_6}
\label{2.6}
\ee

It follows from \rref{2.4} that the temporal components $B^a _{0i}$
are not true dynamical degrees of freedom but rather are Lagrange
multipliers for the constraints
\be 
K^i _a \approx 0 \mbox{.}
\label{2.7}
\ee
These are reducible since
\be 
K^i _{a,i} = 0 \quad \mbox{identically}
\label{2.8}
\ee

Although \rref{2.4} is linear in the time derivatives, it is not yet
in canonical form because the exterior derivative (in field space) of
the one-form $l^{ij}_a \delta B_{ij}^{a}$ is degenerate and thus does
not define a symplectic structure. To deal with this, we
follow the standard Dirac procedure and define the conjugate momenta
through
\be 
p^{ij}_a = \frac{\partial {\cal L}}{\partial \dot{B}^{a}_{ij}} = l^{ij}_a
\label{2.9}
\ee
These momenta are subject to $21N$ primary constraints

\be \Phi ^{ij}_a \equiv p^{ij}_a - l^{ij}_a \approx 0 \quad
i,j=1,\dots
,7;a=1,\dots ,N
\label{2.10}
\ee
$(\Phi ^{ij}_a = - \Phi ^{ji}_a)$.
It turns out to be more convenient to replace the constraints
$K^{i}_a \approx 0$ by the equivalent constraints
\be 
G^{i}_a = K^{i}_a - \partial _j \Phi ^{ij}_a \approx 0
\label{2.11}
\ee
because the new constraints generate the internal gauge
transformations \rref{1.6} in the Poisson brackets,

\be [ B^{a}_{ij}, \int _{\Sigma } d^7 x \Lambda ^b _k G^k _b ] =
( d\Lambda )_{ij}
\label{2.12}
\ee
The constraints \rref{2.11} are also clearly reducible since
$\partial _i G^i _a = 0 $ (identically).

The Hamiltonian action takes the form
\be 
I_{H} = \int dx^0 \int d^7 x [ p^{ij}_a \dot{B}^a _{ij} - B^a
_{0i}
G^i _a
	   - u^a _{ij} \Phi ^{ij}_a ]
\label{2.13}
\ee
The Poisson brackets among the constraints are given by
\begin{eqnarray} 
\left[ {\Phi }^{ij}_a (x) ,{\Phi }^{kl}_b (x') \right]
=&
{\Omega }^{ijkl}_{ab} \delta (x,x' ) \\ 
		  \left[ {\Phi }^{ij}_a (x) , G^l _b \right] =& 0 \\
                 \left[ G^i _a , G^j _b \right] =& 0
\label{2.14}
\end{eqnarray}
where $\Omega ^{ijkl} _{ab}$ is an antisymmetric matrix given by

\begin{eqnarray} 
\Omega ^{ij\ kl} _{a\ b} &=& g_{abc} \epsilon ^{ijklm_1
m_2 m_3} H^c
_{m_1 m_2 m_3} \label{2.15a} \\
		  \Omega ^{ij\ kl} _{a\ b} &=& - \Omega ^{kl\ ij} _{b\ a}
\label{2.15b}
\end{eqnarray}
It follows from the constraint algebra that there are no further
constraints. It is also clear that the
constraints $G^{i} _a \approx 0$ are first class, as they should
since
they generate the internal gauge transformations \rref{1.6}.

\subsection{Rank of $\Omega ^{ijkl}_{ab}$ and projective plane
$\Pi_7$}

\subsubsection{Strategy for computing the rank of $\Omega ^{ijkl}
_{ab}$}

To complete the canonical analysis, it is necessary to determine the
nature of the constraints $\Phi ^{ij}_a \approx 0$. To that end, one
must determine the number of zero eigenvectors possessed by the
matrix
$\Omega ^{ijkl} _{ab}$ of the brackets $[ \Phi ^{ij}_a , \Phi ^{kl}_b
]$. This number turns out to depend on the number $N$ of fields, on
the antisymmetric tensor $g_{abc} $ defining the theory and, for a
given choice of $N$ and $g_{abc}$, it depends also on the phase space
location of the dynamical system. Indeed, the matrix $\Omega ^{ijkl}
_{ab}$ involves both $g_{abc}$ and $H^a _{ijk}$  (the latter being 
constrained by $K^{i} _a \approx 0$).

To determine all the possible ranks that the matrix $\Omega ^{ijkl}
_{ab}$ can achieve is a rather complicated algebraic task and we
shall not attempt to pursue it here. We will  describe
what happens in the {\em generic} situation in which $\Omega ^{ijkl}
_{ab}$
has {\em the maximum possible  rank} compatible with the constraints
$K^{i}
_a \approx 0 $. We call it {generic} because maximum rank
conditions
define open regions in the space of theories (space of the $g$'s) and
of allowed configurations (space of the $H$'s) and are thus stable
under small deformations. This is not true for lower ranks, which are
associated with equations rather than inequalities.

It is easy to see that the matrix $\Omega ^{ijkl} _{ab}$ has at least
seven zero eigenvectors $H^b _{kl\ m} $ ($m=1,2,\dots ,7$), since
$\Omega ^{ijkl} _{ab} H^b _{klm} \approx 0$. These zero eigenvectors
are
associated with the spatial diffeomorphisms in the sense that the
first class constraints
\be
{\cal H}_i \equiv H^a _{ikl} \Phi^{kl}_a \approx 0
\label{2.16}
\ee
are the generators of the spatial diffeomorphisms in the ``improved"
form \rref{1.8}. In the generic case, the zero eigenvectors $H^b
_{klm}$ are independent because the equations
\be
H^a _{ikl} \xi ^i = 0
\label{2.16i}
\ee
imply $\xi ^i = 0$ (more on this in subsection 2.4 below).

The study parallels so far quite closely the discussion of pure
Chern-Simons theories with $1$-form gauge fields. New features arise
when one addresses the question as to whether there are further
independent first class constraints among the $\Phi $ 's. These first
class constraints (if any) would correspond to additional independent
gauge symmetries.

The total number of constraints is the size of $\Omega$, that is to
say $21 N$.  There are as many second class constraints as there are
non-vanishing eigenvalues of $\Omega$.  Therefore, there has to be an
even number of second class constraints.  This means that the number
of first class constraints among the $\Phi$'s is odd for odd $N$ and
even for even $N$.  We already know from (\ref{2.16}) seven of the
first class constraints.  If $N$ is even, there must be at least one
additional first class constraint still to be identified.  This does
not need to be the case for $N$ odd.

By computer-assisted investigation of the
first values of $N$ ($\leq 20$), we have reached the
following conclusions. If $N$ is odd (and greater than or equal to
three), then there is generically no further first class constraint
among the $\Phi $'s; while if $N$ is even (and greater than or equal
to eight in order to avoid the accidental degeneracies), then there
is generically one and only one additional first class constraint
among the $\Phi $'s.

We have come to this conclusion by constructing examples in which
the antisymmetric matrix $\Omega ^{ij\ kl}_{a\ \ b}$ has 
exactly
rank $\kappa = 21N - 7$ for $N$ odd, and exactly rank 
$\kappa - 1 = 21N - 8$ for $N$ even.
Since these values of the rank correspond to the maximum rank that
$\Omega $ can achieve (the first class constraints \rref{2.16} are
always present), they are stable under small deformations and thus
generic.

In order to analyse the equations, we proceed as follows:
\begin{itemize}
\item[i.]
First, we pick at random an arbitrary set of completely antisymmetric
tensors $g_{abc}$;
\item[ii.]
Second, we construct the general solution of the constraints $K^l _a
\equiv g_{abc} H^b _{ijk} H^c _{mnp} \epsilon ^{lijkmnp} = 0$ for
these given $g_{abc}$;
\item[iii.]
Knowing the curvatures $H^a _{ijk}$ (which are subject to the sole
condition $K^l _a \approx 0$ ), we compute the matrix $\Omega ^{ijkl}
_{ab} = \epsilon ^{ijklmnp} g_{abc} H^c _{mnp}$ and determine its
rank.
\end{itemize}

Of these three steps, the first one and the last one are direct. Only
the second one needs further discussion because the equations $K^l _a
= 0$ are quadratic in the unknown $H^b _{ijk}$. It is here that we
shall use the properties of the finite projective plane $\Pi_7$. 

\subsubsection{ Projective plane $\Pi_7$} 

The projective plane $\Pi_7$ has seven points and seven lines
containing three points each,
with the properties that any pair  of points determines one and only
one line, while any pair of lines intersects in one and only one
point. The points and lines of $\Pi_7$ are drawn in figure 2.1, with an
arbitrary choice of indices $(1,2,\dots,N)$.


\begin{center}
\leavevmode
\epsfysize=5cm

\hspace{-.7cm} \epsfbox{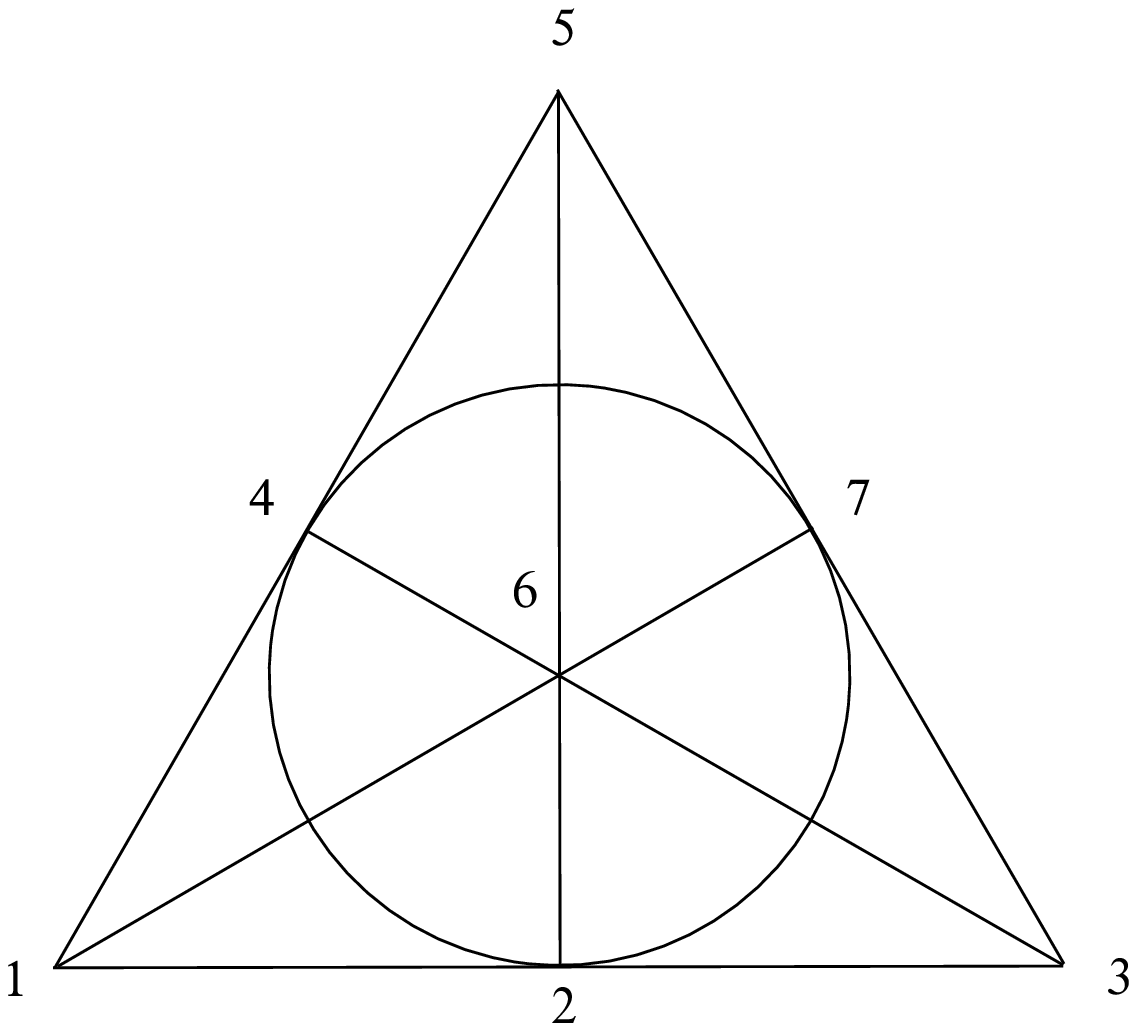}

~\\

 Figure 2.1 \\
\end{center}
{\bf Caption:} The projective plane $\Pi_7$ has seven points
(1,2,3,4,5,6,7) and
seven lines containing three points each.  The
set of these lines is the set of triplets $T = \{ \{1,2,3 \}, \{1,4,5 \},
\{1,6,7\}, \{2,4,7\}, \{2,5,6\}, \{3,4,6\}, \{3,5,7\} \}$.

~\\

To the line $(i,j,k) \in T$, one associates the three-form
$dx^i  \wedge dx^j  \wedge dx^k$.
Because $\Pi_7$ arises in the
description of the octonion algebra (the ``product" of two points is
the third point on the same line, with a sign fixed by the
orientation), we shall call these seven 
3-forms the ``octonionic 3-forms" and shall
denote them by $\omega ^{\alpha } _{(3)} (\alpha = 1,2,\dots ,7)$.
The rest of the basis $3$-forms, formed with triplets not in $T$,
shall be denoted by $\varphi ^A
_{(3)}  (A = 1, \dots , 28)$.  

The main property of the octonionic $3$-forms is
\be 
\omega^{\alpha }_{(3)} \wedge \omega^{\beta}_{(3)} = 0 
\label{2.21}
\ee
for any $\alpha,\beta=(1,2,...,7)$. Indeed, any pair of lines of
$\Pi_7$ have one (and only one if the lines are distinct) point in
common, so that $\omega^{\alpha }_{(3)}$ and $\omega^{\beta}_{(3)}$
have one $dx^i$ in common, which implies \rref{2.21}. 

In what follows we shall use differential form notation in the
spatial manifold. The constraint equations 
\be 
K^i _a = g_{abc} \epsilon ^{ijklmnp} H^b _{jkl} H^c _{mnp} = 0
\label{2.22}
\ee
are rewritten as
\be 
K_a = g_{abc} H^b \wedge H^c = 0 
\label{2.23}
\ee
where the $H^a$ are regarded as three forms in seven dimensions.  
The rank of $\Omega$ is determined by the number of solutions of the
equation
\be
g_{abc} H^b \wedge V^c =0
\label{V}
\ee
where the 3-form curvature $H$ satisfies the constraint equation
(\ref{2.23}) and $V^a$ is a 2-form.  The zero eigenvectors $H^a
_{klm} \xi^m$ described above are simply $i_\xi H^a$.

\subsubsection{The $N=3$ case}

We begin with the $N=3$ case (the smallest number of
fields for which the antisymmetric tensor $g_{abc}$ is non-zero)
because it is particularly straightforward.  A simple solution for
the constraint is given by a linear combination of the octonionic
monomials $H^a=\lambda^a_{\ \alpha} \omega^\alpha_{(3)}$ because
$\omega^\alpha
\wedge \omega^\beta =0$. 

This solution has maximum rank when none of the coefficient
$\lambda^a_{\ \alpha}$ is zero. Indeed, since $H^a$ is a 3-form, its
dual in seven dimensions can be viewed as the symmetric $21\times
21$ matrix $^* H^{(a)\ ij\ kl} = \epsilon^{ijklmnp} H^a_{mnp}$, for
each value of $a=1,2,3$. We can thus write Eq. (\ref{V}) in matrix
form
\be
^*H^1 V^2 = ~^*H^2 V^1,\ \ \ \ ^*H^1 V^3 = ~^*H^3 V^1, \ \ \ \ ^*H^2
V^3 = ~^*H^3 V^2.
\label{eqV}
\ee
The determinant of the matrix $^*H^a$, for a given $a$, is easily
expressed in terms of the expansion $H^a = \lambda^a_\alpha
\omega^\alpha_{(3)}$. Let $^*H^3$ be equal to 
\be
^*H^3 = a_1 ~^*\omega^1 + a_2 ~^*\omega^2 + \cdots + a_7 ~^*\omega^7
\label{H3}
\ee where $^*\omega^\alpha$ represents the dual of $\omega^\alpha$.
The determinant of $^*H^3$ (as a 21$\times$21 matrix) is equal
to $2^7\, a_1 a_2 \cdots a_7$. Thus, if all the monomials $w^\alpha$
are present in (\ref{H3}), $^*H^3$ is invertible. In that case, one
can solve $V^1$ and $V^2$ in terms of $V^3$ from the last two
equations in (\ref{eqV}). Replacing $V^1$ and $V^2$ back in the first
equation we find an equation for $V^3$ \be (^*H^1 J ~^*H^2 - ~^*H^2 J
~^*H^1 ) V^3=0 \label{V3} \ee where $J$ is the inverse of
$^*H^3$. Equation (\ref{V3}) can be investigated numerically and one
finds that it has (generically) only 7 solutions for $V^3$. Since
$V^1$ and $V^2$ are completely determined in terms of $V^3$, this
implies that the matrix $\Omega ^{ijkl}_{ab}$ for the $N=3$ case has
no other zero eigenvalues besides those associated with spatial
diffeomorphisms and has thus maximum rank $21 \times 3 -7 = 56$.

\subsubsection{The general case}

We now consider the general case.  For even $N$, the 
assumption that the $H's$ are
combinations of the octonionic 3-forms does not lead to a maximum
rank for $\Omega$. Thus one needs to look for solutions in which
the
$H's$ have a more general expression. 

Any $3$-form $F$ can be decomposed as
\be 
F = F_{1} + F_{2}
\label{2.17}
\ee
where $F_{1}$ is the ``octonionic part" of $F$,
\be F_{1} = \sum_{\alpha = 1}^7 F_{\alpha } \omega^{\alpha }_{(3)}
\label{2.18}
\ee
and $F_{2}$ is the ``non octonionic part" of $F$,
\be 
F_{2} = \sum_{ A =1}^{28} F_A \varphi ^A _{(3)}
\label{2.19}
\ee

If $G$ is another $3$-form with decomposition $G = G_1 + G_2$, the
exterior product $F \wedge G$ reads 
\be 
F \wedge G = F_1 \wedge G_2 +
F_2 \wedge G_1 + F_2 \wedge G_2 \label{2.20} 
\ee 
There is no term $F_1 \wedge G_1$ because the product of any two
octonionic $3$-form is zero [see (\ref{2.21})]. 

If one decomposes the $3$-forms $H^a$ as in \rref{2.17}, the
constraints (\ref{2.23}) become
\be 
2g_{abc} H^b _1 \wedge H^c _2 + g_{abc} H^b _2 \wedge H^c _2 = 0.
\label{2.24}
\ee 
These equations are actually $7N$ linear equations for the $7N$
octonionic components $H^a _{\alpha }$ of the curvatures. They can be
solved by taking arbitrarily the non octonionic components $H^a _A$ of
the curvatures and determining then the octonionic components $H^a
_{\alpha }$ through \rref{2.24}. [The linear, inhomogeneous system
\rref{2.24} is easily verified to have one and only one solution for
generic $H^a _A $ 's]. Consequently, even though the constraints are
quadratic in $H^a$ one can produce an explicit rational solution by
using the octonionic decomposition.

We have carried out the task of solving the constraint
along the lines just described  for random choices of the
constants $g_{abc}$ and of the components $H^{a}_A$. We have then
computed numerically 
the dimension of the kernel of $\Omega ^{ijkl}_{ab}$ and
found in each case that the zero eigenvalue was degenerate exactly
eight times for $N\ even (\geq 8)$ and seven times for $N\ odd$.
What we got from these calculations for the first
values of $N$ is thus 
that the rank of $\Omega ^{ijkl}_{ab}$ is generically equal
to $21N - 8$ for $N\ even (\geq 8)$ and $21N-7$ for $N\ odd$. 
We have established this result for $N \leq 20$ and shall take it
for granted for higher values of $N$.

\subsection{Geometrical interpretation of the eighth first class
constraint}

 For $N\ odd$ (and $N \geq 3$), the first
class constraints $G^{i}_a \approx 0$ and ${\cal H}_i \approx 0$ are
the only
(independent) first class constraints. The other constraints are
second class. This implies that the internal gauge symmetries
\rref{1.6}, generated by $G^{i}_a$, and the spatial diffeomorphisms,
generated by ${\cal H}_i$, form a complete set of gauge symmetries.
Any gauge symmetry of the system, including the timelike
diffeomorphisms, can be expressed in terms of them.

For $N\ even$ (and $\geq 8$) there is, by contrast, an eighth first
class constraint among the $\Phi ^{ij}_a$ given by $\mu _{ij}^{a}
\Phi ^{ij}_a$ where $\mu _{ij}^{a}$ is a zero eigenvector of $\Omega
^{ijkl}_{ab}$,
\be 
\Omega ^{ijkl}_{ab} \mu ^b_{kl} = 0,
\label{2.25}
\ee
independent of the seven eigenvectors $H^{a}_{ijm}$ associated with
the spatial diffeomorphisms. One may relate the transformation
generated by $\Phi ^{ij}_a \mu^{a}_{ij} \equiv {\cal H}$
\begin{eqnarray} 
\delta _{\rho } B ^{a}_{ij} (\vec{x} ) &=& [ B^a_{ij}
(\vec{x}),
\int \rho (\vec{x}^{\prime }) {\cal H} (\vec{x}^{\prime }) dx^{\prime
} ] \\
	 &=& \rho \mu ^a _{ij} (\vec{x})
\label{2.26}
\end{eqnarray}
to the timelike diffeomorphisms as follows.

In the improved form \rref{1.8}, the timelike diffeomorphisms read

\be \delta _{\xi} B^a _{ij} = \xi ^0 H^a _{0ij}
\label{2.27}
\ee
Now, the equations of motion $g_{abc} H^b \wedge H^c = 0$ imply

\be \Omega ^{ijkl} _{ab} H^b _{0kl} = 0
\label{2.28}
\ee
Thus, $H^b _{0kl}$ is a linear combination of the zero eigenvectors
of 
$\Omega ^{ijkl} _{ab}$,

\be H^a _{0ij} = \lambda \mu ^a _{ij} + \lambda ^k H^a _{kij}
\label{2.29}
\ee
for some $\lambda $ and $\lambda ^k$, which are completely determined
by
$H^a _{0ij}$, $H^a _{ijk}$ and $\mu ^a _{ij}$. As  a consequence, one has
\be 
\delta _{\xi} B^a _{ij} = \xi ^0 \lambda \mu ^a _{ij}
 + \xi ^0 \lambda ^k H^a _{kij}
\label{2.30}
\ee
which shows that the timelike diffeomorphisms can be expressed in
terms of the
spacelike diffeomorphisms and the eighth gauge symmetry \rref{2.26}.
In the generic case, $\lambda$ does not vanish. Accordingly, one can
conversely express the symmetry \rref{2.26} in terms of the spatial
and timelike diffeomorphisms.
\be \mu ^a _{ij} = \frac{1}{\lambda } H^a _{0ij} - \frac{\lambda
^k}{\lambda }
H^a _{kij}
\label{2.31i}
\ee
In that sense, one may say that the eighth first class constraints
present
among the $\Phi $'s when $N$ is even (and $N \geq 8$) generates (a
redefined version of) the timelike diffeomorphisms.

\subsection{Independence of the ${\cal H}_i$'s}

We have asserted above that the ${\cal H}_i$'s defined by \rref{2.16}
were
independent in the generic case, i.e., that for generic case $H^a
_{ijk}$'s
solutions of $K^i _a = 0 $, the system
\be H^a _{ijk} \xi ^{i} = 0
\label{2.31}
\ee
has $\xi ^i = 0$ as only solution. This can be straighforwardly
verified as follows.
It is easy to check that the curvatures automatically solve the
constraints $K^i _a = 0 $ when they have only octonionic components.
This was already used for $N=3$ but clearly holds for
any value of $N$. Furthermore, if for each $\alpha $ there is at
least one value of the index such that $H^a _{\alpha } \neq 0$, then,
the  equations \rref{2.31} imply $\xi^i = 0$. In other words, for
such curvatures, the $21N \times 7$ matrix $H(^a_{ij})_k$ has
maximum rank $7$. Using again the argument that maximum rank
conditions are stable against small deformations, 
one infers that the system $H^a _{ijk} \xi ^i = 0$ implies
generically that $\xi ^i$ is zero.

\subsection{Number of degrees of freedom}

We conclude the discussion of the
$2$-form case by counting the number of
degrees of freedom. There are $21N$ conjugate pairs ($B^a _{ij}$,
$p^{ij}_a$). These pairs are constrained by the $6N$ first class
constraints $G^i _a = 0$ (there are actually $7N$ such constraints,
but they are subject to the differential
identity $G^i _{a,i} = 0$), as well as
by the $21N$ constraints $\Phi ^{ij} _a = 0$. Of these, $7$ are first
class and $21N - 7$ are second class for $N\ odd$ (and $N \geq 3$);
while $8$ are first class and $21N - 8$ are second class for $N\
even$ (and $N \geq 8$). According to the general rule for counting
the degrees of freedom (= number $M$ of physical conjugate pairs),
one finds
\begin{eqnarray*} M &=& \frac{1}{2} [ 2\times 21N - 2\times 6N -
2\times 7 
- (21N - 7) ] \\
      &=& \frac{1}{2} [ 9N - 7 ]  \quad ( N\ odd,\, N \geq 3 )
\end{eqnarray*}
and
\begin{eqnarray*} M &=& \frac{1}{2} [ 2\times 21N - 2\times 6N
-2\times 8 -
 (21N - 8) ] \\
		    &=& \frac{1}{2} [9N - 8 ] \quad (N\ even,\, N \geq 8)
\end{eqnarray*}
For $N = 4$, one has the same number of degrees of freedom as for $N
= 3$, i.e., $10$, while for $N = 6$, one has twice as many degrees of
freedom as for $N = 3$, i.e., $20$. [For $N =1 $ and $2$, the theory
is pure gauge and $M = 0$]. In all cases but $N = 1$ or $2$, the
theory has local degrees of freedom ($M > 0$), as for $1$-forms
\cite{B.G.H.}. This concludes the analysis of the case $p=2,k=2$.

\section{Discussion of the general case ($p \geq 2$, $k\geq 2$)}
\setcounter{equation}{0}

\subsection{Constraints}

We now turn to the general case $p \geq 2$. If $k=1$, the Lagrangian
is quadratic, $ {\cal L} = g_{ab} B^a \wedge H^b $, and the equations
of motion are just equivalent to $H^b = 0$ (assuming $g_{ab}$ to be
invertible). The theory has no local degrees of freedom and one sees
from \rref{1.8} that the diffeomorphisms (both temporal and spatial)
are not independent gauge symmetries. The case $k \geq 2$ is,
however, much richer. Its analysis proceeds as that of the case
$p=2$, $k=2$. We shall use throughout differential form notations.
The number of spatial dimensions is equal to $n-1 = k(p+1)+ p -1$. 

By following the same method as above, one finds that the spatial
field strengths $H^a$ are subject to the algebraic constraints
\be 
K_a^{i_1 \dots i_{p-1}} = g_{ab_1 \dots b_k}
[^*(H^{b_1} \wedge \dots \wedge
H^{b_k})]^{i_1 \dots i_{p-1}} \approx 0,
\label{3.1}
\ee
which can be written as
\be
K_a = g_{a a_1 ... a_k} H^{a_1} \wedge H^{a_2} \wedge ...\wedge
H^{a_k} \approx
0.
\label{cg}
\ee
The $2N \left( \begin{array}{c} n-1 \\ p \end{array} \right)$ phase
space variables $B^a _{i_1 \dots i_p}$ and $p^{i_1 \dots i_p} _a$ are
restricted not only by \rref{3.1}, but also by the primary
constraints.
\be
 \Phi ^{i_1 \dots i_p} _a = p^{i_1 \dots i_p} _a - l^{i_1 \dots
i_p} _a (B) \approx 0
\label{3.2}
\ee
analogous to \rref{2.10}, where $l^{i_1 \dots i_p} _a$ is given by
\be 
l_a^{i_1 \dots i_p} = g_{abc_1 \dots c_{k-1}}[^*(
 B^b\wedge H^{c_1} \wedge\dots\wedge
H^{c_{k-1}})]^{i_1 \dots i_p}. 
\label{3.3}
\ee

There are no further constraints. One can replace \rref{3.1} by
$G^{i_1 \dots i_{p-1}} \approx 0$, where
\be 
G_a^{i_1 \dots i_{p-1}} = K_a^{i_1 \dots i_{p-1}} - 
\partial_i \Phi_a^{i i_1 \dots i_{p-1}} \approx 0.
\label{3.4}
\ee
These constraints generate the internal gauge symmetry \rref{1.6} and
are first class. The constraints \rref{3.2} have brackets given by
\be 
[ {\Phi }_{a}^{i_1 \dots i_p} (\vec{x}), {\Phi }_{b}^{j_1 \dots
j_p} 
     (\vec{x}^{\prime }) ] = {\Omega }^{i_1 \dots i_p j_1 \dots
j_p}_{ab}
      \delta (\vec{x},\vec{x}^{\prime })
\label{3.5}
\ee
where the antisymmetric matrix $\Omega$ is
\be 
\Omega_{ab}^{i_1 \dots i_p j_1 \dots
j_p} = g_{abc_1 \dots c_{k-1}}[^*(H^{c_1} \wedge \dots \wedge
H^{c_{k-1}})]^{i_1 \dots i_p j_1 \dots j_p} 
\label{3.6}
\ee
and satisfies
\be
\Omega^{i_1 \dots i_p j_1 \dots j_p} _{ab} =
   - \Omega ^{j_1 \dots j_p i_1 \dots i_p} _{ba}. 
\label{3.7}
\ee

The number of independent gauge symmetries is equal to the number of
independent first class constraints. To determine this number, we have
to find how the constraints $\Phi ^{i_1 \dots i_p}$ split into first
class constraints and second class constraints. To that end, we must
compute the number of zero eigenvalues of $\Omega ^{i_1 \dots i_p j_1
\dots j_p} _{ab}$.

Now, just as in the $p=2$, $k=2$ case of the previous section, the 
$H^b _{j_1 \dots j_p l}$ are $n-1$ zero eigenvalues of 
$\Omega ^{i_1 \dots i_p j_1 \dots j_p} _{ab}$,
\be
\Omega ^{i_1 \dots i_p j_1 \dots j_p} _{ab} H^b _{j_1 \dots j_p
l} = 0
\label{3.8}
\ee
This is direct consequence of \rref{3.1} and is most easily
verified observing that the zero eigenvalues of $\Omega$ can be
viewed as $p$-forms $\mu^a$ satisfying 
\begin{equation}
g_{ab_1...b_k} H^{b_1} \wedge ...\wedge H^{b_{k-1}} \wedge \mu^{b_k}
=0 .
\label{zero/eigen}
\end{equation} 
Since $i_\xi$ is an antiderivation, one obtains from (\ref{cg}) that 
$i_\xi H^a$ is a solution of (\ref{zero/eigen}) for any spatial
vector $\xi^i$.

The corresponding first class constraints
\be 
{\cal H} _l = H^b _{j_1 \dots j_p l} \Phi ^{j_1 \dots j_p} _b
\label{3.9}
\ee
generate the spatial diffeomorphisms. The question is: Are
there further first class constraints among the $\Phi $'s ?
This depends on whether the number
$$
\kappa  \equiv N \left( \begin{array}{c} n-1 \\ p
\end{array} \right) - (n-1)  
$$
is even or odd (we remind that $ n = k(p+1) + p $). 

The antisymmetric matrix $\Omega ^{i_1 \dots i_p j_1 \dots j_p}
_{ab}$ is necessarily of even rank. Accordingly, if $\kappa$ is even,
there is no a priori reason that the rank of $\Omega$, known to be
$\leq \kappa$, could not be precisely equal to $\kappa$.  
By constrast, if $\kappa$
is odd, there is definitely one extra zero eigenvector implying the
existence of a further independent gauge symmetry, and the rank of
$\Omega$ is at most equal to $\kappa-1$.

In the generic case, the rank of $\Omega$ is expected to be equal to the
maximum value compatible with the existence of the known zero
eigenvalues $i_\xi H$, i.e. $\kappa$ ($\kappa$ even) or $\kappa-1$ 
($\kappa$ odd).   This is supported by the results found
for $p=1$ \cite{B.G.H.}, $p=2$, $n=8$ (see above) and
$p=3$, $n=11$, $N=1$ (section 3.3. below).  Thus,
there will be no further independent gauge symmetries besides the
internal gauge symmetries and the spatial diffeomorphisms ($\kappa$ even),
or the internal gauge symmetries, the spatial diffeomorphisms and the
timelike diffeomorphisms ($\kappa$ odd). Other independent gauge
symmetries could arise for particular choices of field configurations
or tensor $g_{a_1...a_{k+1}}$, but these should be thought of as
accidental. 

Of course, the above comments would be somewhat empty if $\kappa$ was
always even. But we have seen that at least for $p=2$, one may have
an odd $\kappa$. This is not the only case. The number $\kappa$ may be odd
whenever $p$ is even. To prove this statement, we must investigate
the parity
of $\kappa$ as a function of the form-degree $p$, the number $N$ of fields
and the order $k$ of the Chern-Simons theory.  

\subsection{Case $p$ odd}

If the form degree $p$ is odd, the spatial dimension $n-1 = k(p+1)
+p-1$ is even
since both $p+1$ and $p-1 $ are even. It is straightforward
to see that in a
space with an even number of dimensions, a form of odd degree has an
even number of components,
\be 
\left( \begin{array}{c} u \\ p \end{array} \right) = \mbox{even
number}
\quad \mbox{( u even, p odd)}
\label{3.10}
\ee
$(u=n-1)$. More precisely,
\begin{eqnarray} 
\left( \begin{array}{c} u \\ p \end{array} \right)
=&  {\displaystyle {u(u-1)(u-2)\dots (u-p+1)}}
\over{\displaystyle{1. 2 . 3  \dots p}} \\
=& {\displaystyle u\; K}\over{\displaystyle p}
\label{3.11}
\end{eqnarray}
where $K$ is the integer 
\begin{eqnarray*}
K = \left( \begin{array}{c} u-1 \\ p-1 \end{array} \right) 
= \frac{(u-1)(u-2) \dots (u-p+1)}{1. 2 .3 \dots   (p-1)}.
\end{eqnarray*}
Since the left-hand side of (\ref{3.11}) is an integer, $u K$ is a
multiple of $p$.  Since $u$ is even, the product $u K$ is even.
Consequently, $u K$ is a multiple of $2p$ since $p$ is odd.  Hence
${u.K}/{p}$ is even.

It follows that the integer $\kappa$, given by the difference between
two even integers, is also an even integer. Thus, whenever $p$ is odd,
the rank of $\Omega$ should be exactly equal to $\kappa$ (barring
accidental degeneracies) and the temporal diffeomorphisms should not
be independent gauge symmetries. This is exactly as in the $p=1$ case
studied in \cite{B.G.H.}.

\subsection{The eleven dimensional $H\wedge H\wedge B$ theory}

Another interesting illustration of the results just stated is given
by the case $N=1$, $p=3$ and $k=2$ (implying $n=11$) for which the
Chern-Simons action reads
\begin{equation}
I= \int H \wedge H\wedge B
\label{I/11}
\end{equation}
This term arises in supergravity in eleven
dimensions \cite{julia} which has recently received much 
attention in the context of $M-$ theory. 

To show that $\Omega$ has generically maximum rank 120 (number of
primary constraints) - 10 (number of spatial diffeomorphisms) = 110,
it is enough to exhibit one 4-form $H$, solution of the constraint
equation
\begin{equation}
H \wedge H=0,
\label{c/11}
\end{equation}
for which this property is fulfilled. 

A solution for (\ref{c/11}) with maximum rank can be constructed as
follows.  We consider the following  4-form,
\begin{eqnarray}
H =  ~&& f\ dx^1 \wedge dx^3 \wedge dx^5 \wedge dx^7 +  
      2  dx^2 \wedge dx^4 \wedge dx^6 \wedge dx^8   \nonumber \\
&-&\alpha\ dx^1 \wedge dx^2 \wedge dx^3 \wedge dx^4 -
\beta\  dx^1 \wedge dx^2 \wedge dx^5 \wedge dx^6  \nonumber \\
&+&      a\ dx^1 \wedge dx^2 \wedge dx^7 \wedge dx^8  + 
 b\      dx^1 \wedge dx^2 \wedge dx^9 \wedge dx^{10}   \nonumber\\ 
&-&\gamma\ dx^3 \wedge dx^4 \wedge dx^5 \wedge dx^6  + 
c\      dx^3 \wedge dx^4 \wedge dx^7 \wedge dx^8  \nonumber\\
&+&d\       dx^3 \wedge dx^4 \wedge dx^9 \wedge dx^{10} +
A\    dx^5 \wedge dx^6 \wedge dx^7 \wedge dx^8 \nonumber \\  
&+&B\     dx^5 \wedge dx^6 \wedge dx^9 \wedge dx^{10} +
       dx^7 \wedge dx^8 \wedge dx^9 \wedge dx^{10} \nonumber 
\end{eqnarray}
where $a,b,c,d,\alpha,\beta,\gamma,A,B,f$ are  constants. This
form of $H$ has been found by trial and error, by successive
complications of the initial attempt that involved only products of
the binomials $dx^1\wedge dx^2, dx^3\wedge dx^4, dx^5\wedge dx^6,
dx^7\wedge dx^8, dx^9\wedge dx^{10}$. 

The equation $H \wedge H=0$ imposes the  following restrictions among
the coefficients
\begin{eqnarray}
\alpha A + \beta c + \gamma a=2f \label{1}  \\
\alpha B + \beta d + \gamma b=0  \label{2}
\end{eqnarray}
and
\begin{eqnarray}
\alpha = ad +bc \nonumber\\
\beta = a B + b A \label{alpha}\\
\gamma=c B + d A \nonumber
\end{eqnarray}
Replacing (\ref{alpha}) in (\ref{1}) and (\ref{2}) we obtain a system
of equations for $A$ and $B$ whose solution is
\begin{equation}
A= - \frac{(ad+bc)f}{acbd -(ad+bc)^2}, \ \ \ \   
B=\frac{bdf}{acbd-(ad+bc)^2}
\end{equation}
Therefore, $\alpha,\beta,\gamma$ and $A,B$ are determined in terms of
the parameters $a,b,c,d$ and $f$, which are left arbitrary in the
solution. 

We have computed the rank of $\Omega$ for generic values of the
coefficients $a,b,c,d$ and $f$ and found that the maximum rank is
achieved.  The $120\times 120$ matrix $\Omega$ has only
10 zero eigenvalues which correspond to the 10 independent
diffeomorphisms of the spatial surface.   Note that this theory has
19 local degrees of freedom, as easily checked by using the rule for
counting degrees of freedom recalled above. 

\subsection{ Case $p$ even}

If the form degree $p$ is even, $\kappa$ can be even or odd.
As a function of the number of fields $N$ for fixed $p$ and $k$,
various
possibilities may actually arise:

\begin{itemize}
\item[i)] $\kappa$ is even no matter $N$ is;
\item[ii)] $\kappa$ is even for odd $N$ and odd for even $N$;
\item[iii)] $\kappa$ is even for odd $N$ and odd for odd $N$;
\item[iv)] $\kappa$ is odd no matter what $N$ is.
\end{itemize}

We have found an instance of (ii) in the previous case $p=2$,
$k=2$. To illustrate the other possibilities, we just consider other
values of $k$ while keeping $p$ fixed and equal to two.

\begin{itemize}
\item[i)]
$k = 5$: the number of spatial dimensions is equal to
$5\times 3 +1 = 16$. A $2$-form has 
$\left( \begin{array}{c} 16 \\ 2 \end{array} \right) = \frac{16\times
15}{2} = 120$
spatial components. Thus $\kappa = 120N - 16$ is even no matter what the
integer
$N$ is.
\item[iii)]
$k=3$: $n-1 = 10$ and $\left( \begin{array}{c} n-1 \\ 2 \end{array}
\right) = 45$.
Thus $\kappa = 45 N - 10$ is even for $N$ even and odd for $N$ odd.
\item[iv)]
$k=4$: $n-1 = 13$ and $\left( \begin{array}{c} n-1 \\ 2 \end{array}
\right) = 78$. Thus $\kappa = 78N - 13$ is odd no matter what the integer
$N$ is.

\end{itemize}

\section{Conclusions}
\setcounter{equation}{0}

We have established that the nature of the independent gauge
symmetries of pure Chern-Simons theory based on $p$-form gauge fields
crucially depends on the parity of $p$. While the spacelike
diffeomorphisms and the internal gauge symmetries are the only
independent gauge symmetries for odd $p$ (in the absence of,
non-generic, extra gauge symmetries), the timelike diffeomorphisms are
independent gauge symmetries when $p$ is even, if, in addition, the
number $ \kappa\equiv N \left(\begin{array}{c} n-1 \\ p \end{array}
\right) -(n-1)$ is an odd integer.  The difference between odd $p$ and
even $p$ persists even in the presence of accidental gauge symmetries,
which must necessarily come in pairs since the rank of $\Omega$ is
necessarily even.  For odd values of $p$, the timelike diffeomorphisms
can be expressed in terms of the spacelike diffeomorphisms, the
internal gauge symmetries and the (even number of) accidental gauge
symmetries.  This is not true for even $p$ (and odd $\kappa$).

This result is somehow unexpected since the construction of the
Chern-Simons Lagrangian is identical in all cases. Furthermore, for
fixed (even) $p$ and fixed spacetime dimension $n$, the parity of
$\kappa$ may depend - again somewhat surprisingly - on the number $N$
of $p$-forms involved. We have not provided a deep explanation of why
the distinction between the cases of even or odd $p$ arises, but we
have provided explicit examples illustrating both situations.  Perhaps
an investigation of Chern-Simons theories involving simultaneously
forms of different degrees could shed further light on the question.

\section{Acknowledgements}

MB was partially supported by grants \# 1960065 and \# 1970150 from
FONDECYT (Chile), and institutional support by a group of Chilean
companies (EMPRESAS CMPC, CGE, COPEC, CODELCO, MINERA LA
ESCONDIDA, NOVAGAS, ENERSIS, BUSINESS DESIGN ASS. and XEROX Chile).

\end{document}